\documentclass[11pt]{article}
\usepackage{epsfig}
\usepackage{latexsym}
\usepackage{amsmath}
\usepackage[noadjust]{cite}
\usepackage [mathlines, displaymath]{}
\title {{\bf Long-term evolution of an ecosystem with spontaneous periodicity of mass
extinctions}}
\author{{\sc Adam Lipowski}$^1$\ \ and {\sc Dorota Lipowska}$^2$\\
\noalign{\vskip5mm}
$^1$Faculty of Physics, Adam Mickiewicz University,\\
61-614 Pozna\'{n},
Poland\\
$^2$Institute of Linguistics, Adam Mickiewicz University,\\
60-371 Pozna\'{n}, Poland}
\date{}
\newcommand{\cyt} [1] {\cite{#1}}
\def\bib#1#2\par{\bibitem{#1} #2 \bibetykieta{#1}}
\begin {document}

%\renewcommand\@biblabel[1]{#1.}
%\makeatletter \renewcommand\@biblabel[1]{#1.} \makeatother  % numerek z kropka
\makeatletter \renewcommand\@biblabel[1]{} \makeatother

%%%%%%%%%%%%%%%%%%%%%%%%%%%%%%%%%%%%%%%%%%%%%%%%%%%%%%%%%
\maketitle

%Running title: Long-term evolution of an ecosystem\\

%\linenumbers
\begin{abstract}
Twenty years ago, after analysing palaeontological data, Raup and
Sepkoski suggested that mass extinctions on Earth appear
cyclically in time with a period of approximately 26 million years
(My). To explain the 26My period, a number of proposals were made
involving, e.g., astronomical effects, increased volcanic
activity, or the Earth's magnetic field reversal, none of which,
however, has been confirmed. Here we study a spatially extended
discrete model of an ecosystem and show that the periodicity of
mass extinctions might be a natural feature of the ecosystem's
dynamics and not the result of a periodic external perturbation.
In our model, periodic changes of the diversity of an ecosystem
and some of its other characteristics are induced by the
coevolution of species. In agreement with some palaeontological
data, our results show that the longevity of a species depends on
the evolutionary stage at which the species is created. Possible
further tests of our model are also discussed.
\end{abstract}
\section{Introduction}
Recently, numerous efforts are being made to describe the
large-scale evolution of the Earth ecosystem. Due to the
ecosystem's complexity, however, such a task is extremely
difficult~(\cyt{pimm}). That is why researchers in this field have
to turn to very simplified and abstract models that hopefully
still contain relevant factors. Such an approach proved to be
successful, e.g., in modelling of some aspects of extinction
dynamics (\cyt{NEWMAN}). Indeed, in very simple models of
ecosystems certain properties of extinctions as, e.g., the
distribution of sizes or durations of extinctions, seem to agree,
at least qualitatively, with palaeontological data
(\cyt{BAKSNEPP}; \cyt{SOLE}). In these models, the dynamics
spontaneously drives the ecosystem toward the scale-invariant
state with extinctions described by some power-law
characteristics. However, since the accuracy of fossil data is
rather limited, especially with respect to events on a large
timescale, the applicability of such models should be considered
with care.

The suggestion that the extinction dynamics is not scale invariant
but it has a characteristic timescale was made by Raup and
Sepkoski (\cyt{raupsep}). While analysing fossil data, they
noticed that during the last 250 My mass extinctions on Earth
appeared more or less cyclically with a period of approximately
26My. Although their analysis was initially questioned
(\cyt{patterson}), some other works confirmed Raup and Sepkoski's
hypothesis (\cyt{fox}; \cyt{prokoph}; \cyt{plotnick}). The
suggested large periodicity of mass extinctions turned out to be
very difficult to explain. Indeed, 26My does not seem to match any
of known Earth cycles and some researchers have been looking for
more exotic explanations involving astronomical effects
(\cyt{theories1}; \cyt{theories2}), increased volcanic activity
(\cyt{stot1}), or the Earth's magnetic field reversal
(\cyt{stot2}). So far, however, none of these proposals has been
confirmed. One should also note that the most recent analysis of
palaeontological data that span last 542My strongly supports the
periodicity of mass extinctions albeit with a larger cycle of
about 62My (\cyt{rohde}).
%Moreover, we examine universality in the time dependence of the size of a
%population of species.

%%%%%%%%%%%%%%%%%%%%%%%%%%%%%%%%%%%%%%%%%%%%%%%%%%%%%%%%%

Lacking a firm evidence of an exogenous cause, one can ask whether
the periodicity of extinctions can be explained without referring
to such a factor. It is already well known that a periodic
behaviour of a system is not necessarily the result of periodic
driving. In particular, since the seminal works of Lotka and
Volterra, it is known that spontaneous oscillations of the
population size might appear in various prey-predator systems~
(\cyt{LV}). However, the period of oscillations in such systems is
determined by the growth and death rate coefficients of
interacting species and is of the order of a few years rather than
tens of millions. Consequently, if the periodicity of mass
extinctions is to be explained within a model of interacting
species, a different mechanism that generates long-period
oscillations must be at work.

Recently, a multi-species prey-predator model has been introduced,
where long-term oscillatory behaviour is observed
(\cyt{lipowski}). Only some preliminary studies of basic
properties of this model have been made, and the objective of the
present paper is to provide its more detailed analysis. In this
model the period of oscillations is determined by the inverse of
the mutation rate and as we argue, it should be several orders of
magnitude longer than in the Lotka-Volterra oscillations. The
mechanism that generates oscillations in our model can be briefly
described as follows: A coevolution of predator species induced by
the competition for food and space causes a gradual increase of
their size. However, such an increase leads to the overpopulation
of large predators and a shortage of preys. It is then followed by
a depletion of large species and a subsequent return to the
multi-species stage with mainly small species that again gradually
increase their size and the cycle repeats. Numerical calculations
for our model show that the longevity of a species depends on the
evolutionary stage at which the species is created. A similar
pattern  has been observed in some palaeontological data
(\cyt{aimiller}) and, to our knowledge, the presented model is the
first one that reproduces such a dependence. Let us notice that
the oscillatory behaviour in a prey-predator system that was also
attributed to the coevolution has been already examined by
Dieckmann et al.~(\cyt{DIECKMANN}). In their model, however, the
number of species is kept constant and it cannot be applied to
study extinctions. Moreover, the idea that an internal ecosystem
dynamics might be partially responsible for the long-term
periodicity in the fossil records was suggested by
Stanley~(\cyt{STANLEY}) and later examined by Plotnick and
McKinney (\cyt{plotnick1993}). However, in his approach mass
extinctions are triggered by external impacts. Their approximately
equidistant separation is the result of a delayed recovery of the
ecosystem. In our approach no external factor is needed to trigger
such extinctions and sustain their approximate periodicity.
%Although our model is only a drastic simplification of real
%ecosystem, in our opinion it takes into account its important
%characteristics (...)
\section{Model}
Numerical simulations and models of various levels of description
have been frequently used to study extinctions of species
(\cyt{NEWMAN}). In the simplest cases, the dynamics of models was
formulated at the level of species and had to refer to the notion
of fitness that is not commonly accepted. In more recent
approaches, an individual-oriented dynamics has often been used
and although computationally more demanding, such models are
considered as more adequate (\cyt{higgs}; \cyt{stauffer};
\cyt{fdl}). Our model uses the individual-oriented dynamics but in
addition it is spatially extended. Such a feature increases the
computational complexity even more but it also takes into account,
e.g., dynamically generated spatial inhomogeneities that sometimes
are known to play an important role. Our model can be also
considered as a multi-species generalization of the already
studied spatially extended prey-predator model (\cyt{lip1999};
\cyt{lip2000}). Some other multi-species lattice models were also
studied in various contexts (\cyt{pekal}; \cyt{SATO};
\cyt{DIECKMANN2000}).

Our model describes a multi-species prey-predator system defined
on a square lattice of linear size $N$ (\cyt{lipowski}). At each
site of a lattice $i$ there is an operator $x_i$ that specifies
whether this site is occupied by a prey ($x_i=1$), by a predator
($x_i=2$), by both of them ($x_i=3$), or is empty ($x_i=0$). Each
predator is characterized by its size $m \ (0<m<1)$ that
determines its consumption rate and at the same time its strength
when it competes with other predators. Only approximately the size
$m$ can be considered as related with physical size. Predators and
preys evolve according to rules typical to such systems (e.g.,
predators must eat preys to survive, preys and predators can breed
provided that there is an empty site nearby, etc.). In addition,
the relative update rate for preys and predators is specified by
the parameter $r \ (0<r<1)$ and during breeding mutations are
taking place with the probability $p$. More detailed definition of
the model dynamics is given below:\\
(a) Choose a site at random (the chosen site is denoted by $i$).\\
(b) Provided that $i$ is occupied by a prey (i.e., if $x_i=1$ or
$x_i=3$) update the prey with the probability $r$. If at least one
neighbor (say $j$) of the chosen site is not occupied by a prey
(i.e., $x_j=0$ or $x_j=2$), the prey at the site $i$ produces an
offspring and places it on an empty neighboring site (if there are
more empty sites, one of them is chosen randomly). Otherwise
(i.e., if there are no empty sites) the prey does not breed.\\
(c) Provided that $i$ is occupied by a predator (i.e., $x_i=2$ or
$x_i=3$) update the predator with the probability $(1-r)m_i$,
where $m_i$ is the size of the predator at site $i$. If the chosen
site $i$ is occupied by a predator only ($x_i=2$), it dies, i.e.,
the site becomes empty ($x_i=0$). If there is also a prey there
($x_i=3$), the predator consumes the prey (i.e., $x_i$ is set to
2) and if possible, it places an offspring at an empty neighboring
site. For a predator of the size $m_i$ it is possible to place an
offspring at the site $j$ provided that $j$ is not occupied by a
predator ($x_j=0$ or $x_j=1$) or is occupied by a predator
($x_j=2$ or $x_j=3$) but of a smaller size than $m_i$ (in such a
case the smaller-size predator is replaced by an offspring of the
larger-size predator). The offspring inherits its parent's size
with the probability $1-p$ and
with the probability $p$ it gets a new size that is drawn from a uniform distribution.\\

At first sight one can think that such a model describes an
ecosystem with two trophic levels (preys and predators) and only
with predators being equipped with evolutionary abilities, which
would be of course highly unrealistic. Let us notice, however,
that expansion of predators sometimes proceeds at the expense of
smaller-size predators. Thus, predators themselves  are involved
in prey-predator-like interactions. Perhaps it would be more
appropriate to consider unmutable preys as a renewable (at a
finite rate) source of, e.g., energy, and predators as actual
species involved in various prey-predator interactions and
equipped with evolutionary abilities.
%%%%%%%%%%%%%%%%%%%%%%%%%%%%%%%%%%%%%%%%%%%%%%%%%%%%%%%%%%%%%%%%%%%%%%
\section{Results}
To examine the behaviour of this model we used numerical
simulations. Our results, shown in
Figs.\ref{okna}-\ref{spsize}, are obtained for $r=0.2$ but we
expect (\cyt{lipowski}) that the behaviour of the model should be
qualitatively the same for any $r<0.27$ (a brief discussion of the
behaviour of the model for $r>0.27$ is given at the end of this
section).
\subsection{Oscillatory behaviour}
\begin{figure}
%\vspace{-3cm}
\vspace{-0cm} \centerline{ \epsfxsize=9cm \epsfbox{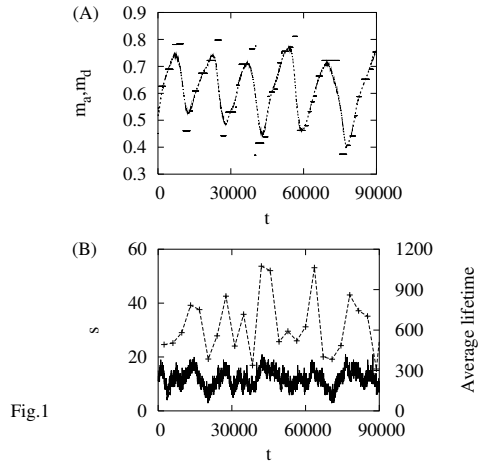} }
%\figspace
\caption{The results of numerical simulations ($N=500$,\
p=0.00001). Data on both panels are obtained from the same run. In
our simulations a unit of time is defined as a single on average
update of each site. (A) The time dependence of the average size
$m_a$ (dashed line) and the size of the dominant species $m_d$
(short, horizontal intervals). (B) The number of species $s$
(continuous line) and the averaged lifetime of species (+). After
extinction the number of species drops 3-4 times. To reduce
stochastic noise in the calculation of the average lifetime data
are collected in time windows of the width $\Delta t = 3000$.}
\label{okna}
\end{figure}

In Fig.\ref{okna}B one can see that, indeed, the number of species
$s$ exhibits pronounced irregular oscillations. These oscillations
are coupled with more regular oscillations of the averaged (over
all predators) size $m_a$ (Fig.\ref{okna}A) and maxima of $s$
correspond approximately to minima of $m_a$ and vice versa. To
have a better understanding of the behaviour of the model we also
calculated the size $m_d$ of the dominant species (i.e., the
predator species with the largest number of individuals) and the
results are shown in Fig.\ref{okna}A.
%%%%%%%%%%%%%%%%%%%%%%%%%%%%%%%%%%%%%%%%%%%%%%%%%%%%%%%%%%%%
%%%%%%%%%%%%%%%%%%%%%%%%%%%%%%%%%%%%%%%%%%%%%%%%%%%%%%%%%%%%

\begin{figure}
%\vspace{-3cm}
\vspace{0cm} \centerline{ \epsfxsize=9cm \epsfbox{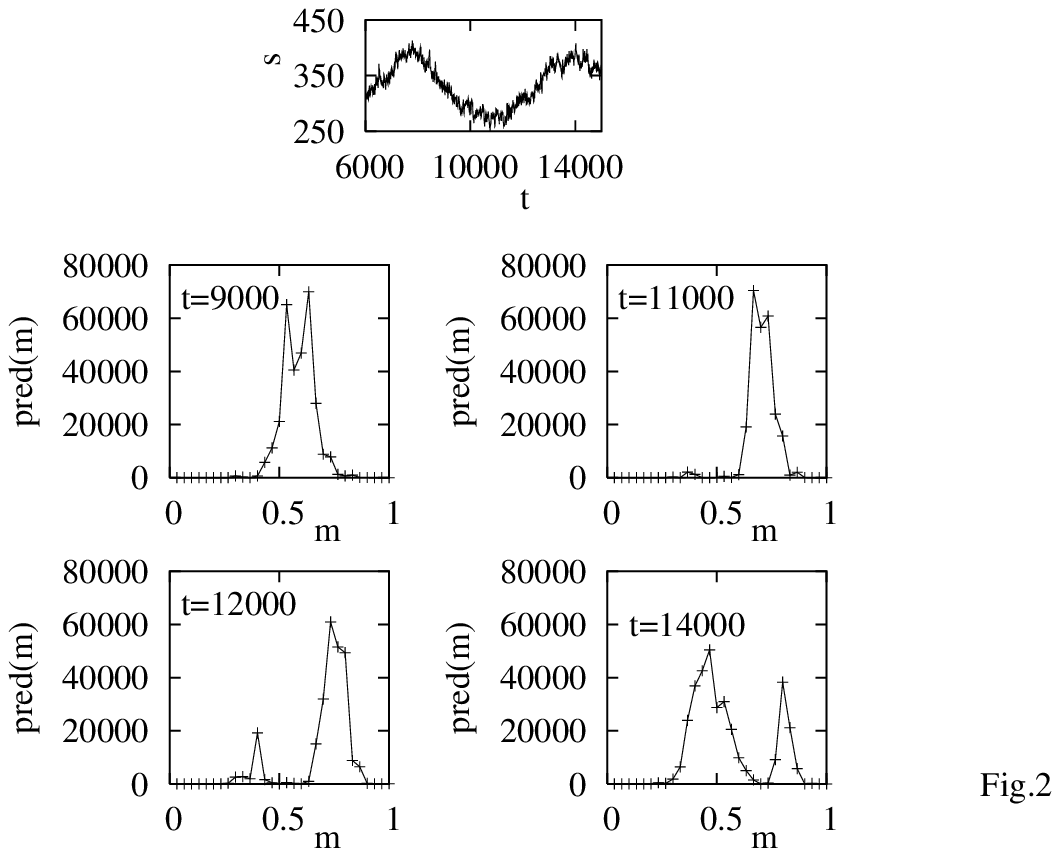} }
%\figspace
\caption{Four panels show distribution of sizes of species at
various stages of evolutionary cycle ($N=1000$,\ p=0.0001). The
upper panel shows the time dependence of the number of species
$s$.} \label{sizes}
\end{figure}

These results indicate that the behaviour of our model can be
described as follows: In a species-rich interval the size $m_a$ is
typically quite low and there is an abundance of preys. In such a
case predators of a large size are in a more favorable position
(because a larger predator can replace a smaller predator) and as
a result $m_a$ and $m_d$ increase. The process of increasing the
size is gradual and involves a large number of species and is not
related to a creation of a single (very-efficient) species, as
suggested previously (\cyt{lipowski}). The increased size $m$
implies a higher consumption rate and due to a finite recovery
rate of preys the large-size species, that at this stage dominate
the system, are running out of food. At first sight one might
expect that further evolution will gradually reduce $m_a$ and
$m_d$. Numerical results show, however (Fig.\ref{okna}A), that
after reaching a local maximum, $m_d$ jumps to a very low value.
This indicates that abrupt changes take place in the model after
which large-size species are no longer dominant and vast majority
of them become extinct. At the same time, however, a lot of new,
mainly small-size species is created and that increases the
diversity $s$ (although we do not suggest that this was really the
cause, a succession of small mammals after large dinosaurs could
be a vivid example of such a change). In such a way the system
returns to the initial species-rich state. Such a cycle is also
illustrated in Fig.\ref{sizes} that shows the distribution of size
$m$ at various stages of the evolution\footnote{ Dynamics of the
model is also illustrated with a Java applet available at: http://
spin.amu.edu.pl/lipowski/prey\_pred.html}.

%%%%%%%%%%%%%%%%%%%%%%%%%%%%%%%%%%%%%%%%%%%%%%%%%%%%%%%%%%%%
%%%%%%%%%%%%%%%%%%%%%%%%%%%%%%%%%%%%%%%%%%%%%%%%%%%%%%%%%%%%

Gradual increase of size of species recalls the Cope's rule that
states that species tend to increase body size over geological
time. This rule is not commonly accepted among paleontologists and
evolutionists and was questioned on various grounds
(\cyt{STANLEY1973}). However, recent studies of fossil records of
mammal species are consistent with this rule (\cyt{ALROY};
\cyt{VAN}). Perhaps our model could suggests a way to obtain a
theoretical justification of this rule.

From the above description, it is expected that the periodicity of
such a cycle increases when the mutation rate $p$ decreases, and
such a behaviour is confirmed with  more detailed calculations
(\cyt{lipowski}). In particular, already for $p=10^{-5}$ the
estimated (\cyt{lipowski}) periodicity of oscillations in our
model is approximately 1000 times larger than that of the
Lotka-Volterra oscillations in the corresponding single-predator
system. It shows that the oscillations in our model are indeed
long-period and, perhaps for smaller $p$, on a timescale close to
26My.

Although very complicated, in principle, it should be possible to
estimate the value of the mutation probability $p$ from the
mutational properties of living species. Let us notice that in our
model mutations produce an individual that might be substantially
different from its parent. In Nature, this is typically the result
of many cumulative mutations and thus we expect that $p$ is indeed
a very small quantity. Actually, $p$ should be considered rather
as a parameter related with the speed of morphological and
speciation processes that are known to be typically very slow
(\cyt{gingerich}). Perhaps a modification of the mutation
mechanism where a new species will be only a small modification of
its parental species could be more suitable for comparison with
living species, but it might require longer calculations.
Alternatively, one can try to estimate the parameters $p$, $N$,
and $r$ (or at least their ratios) by matching the behaviour of
our model with some characteristics of the ecosystem such as the
period of oscillations (26My), fraction of extinct species during
a mass extinction or the average lifetime of species as compared
with the periodicity of mass extinctions.

The oscillatory behaviour sets probably the largest timescale in our
model. However, on the shorter timescale some characteristics,
such as, e.g., the number of species, exhibit strong fluctuations
(Fig.~\ref{okna}B). On such a timescale some distributions might
be very broad and resemble power-law distributions. Indeed, such a
behaviour was demonstrated for the distribution of lifetimes of
species in our model (\cyt{lipowski}).

The increase of the size of species in our model resembles the
fitness-increasing evolution in the real ecosystem. It is tempting
to consider present-day large mammals as highly adapted dominant
species and, in the context of our model, located perhaps close to
the local maximum in the fitness space (as in Fig.\ref{okna}A). If
so, then according to our model, the next dominant species most
likely will be a small-size species that at the moment might not
even exist. Its dominance will be possible due to drastic and
inevitable changes of our ecosystem. Putting aside the validity of
our model, such a scenario does not seem unlikely.

\subsection{Longevity of species}
An analysis of palaeontological data (\cyt{aimiller}) shows that
the longevity is larger for species created after mass extinctions
than for other species. To compare such a result with the
predictions of our model, we calculated the average lifetime of
species. It turns out, however, that important contributions to
this quantity are coming from short-living species and their
lifetime is essentially independent on the evolutionary phase at
which they are born. To reduce this effect we took into account
only the species that lived longer than a given threshold, which
we set equal to 30. Fossils of species of short lifetime are
rather scarce and palaeontological data also reflect a similar
bias toward long-lifetime species. The obtained results are shown
in Fig.\ref{okna}B. Although still strongly fluctuating, they
clearly show that the lifetime is correlated with the global
evolution of the ecosystem and they qualitatively agree with
palaeontological data. In particular, the maximum lifetime appears
for species born shortly after a large and abrupt decrease of the
size of the dominant species (crash). Apparently, species created
at this time find most favourable conditions while the worst
conditions exist shortly before a crash. Again using the analogy
with the real ecosystem and humans, the model predicts (not
counter-intuitively) that species created during our dominance
will have a rather short lifetime.

%%%%%%%%%%%%%%%%%%%%%%%%%%%%%%%%%%%%%%%%%%%%%%%%%%%%%%%%%%%%
\begin{figure}
%\vspace{-3cm}
\vspace{0cm} \centerline{ \epsfxsize=9cm\epsfbox{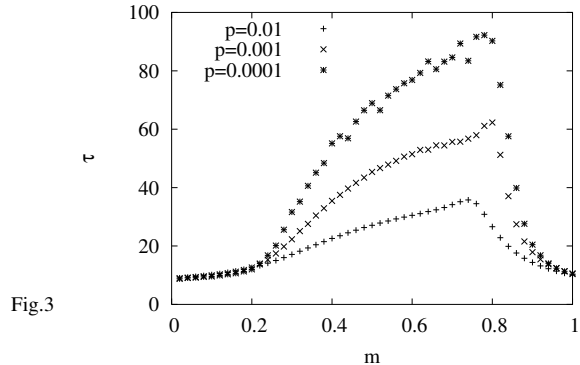}
}
%\figspace
\caption{The average lifetime of species $\tau$ as a function of
the size $m$ ($N=500$).} \label{rate_tau.eps}
\end{figure}
%%%%%%%%%%%%%%%%%%%%%%%%%%%%%%%%%%%%%%%%%%%%%%%%%%%%%%%%%%%%
%%%%%%%%%%%%%%%%%%%%%%%%%%%%%%%%%%%%%%%%%%%%%%%%%%%%%%%%%%%

For a species to have a very small size $m$ is usually a
disadvantage since such a species will loose in competition with
other species. On the other hand, a large size implies a high
consumption rate and such a species might suffer from lack of
food. It means that a lifetime of a species as a function of $m$
should  have a maximum at a certain intermediate value and
numerical calculations confirm such a behaviour (see
Fig~\ref{rate_tau.eps}). Some data on distribution of sizes in
Pleistocene and Recent molluscan faunas do show some maximum
(\cyt{jablonsky}) but a more detailed comparison cannot be done
yet.

\begin{figure}
%\vspace{-3cm}
\vspace{0cm} \centerline{ \epsfxsize=9cm \epsfbox{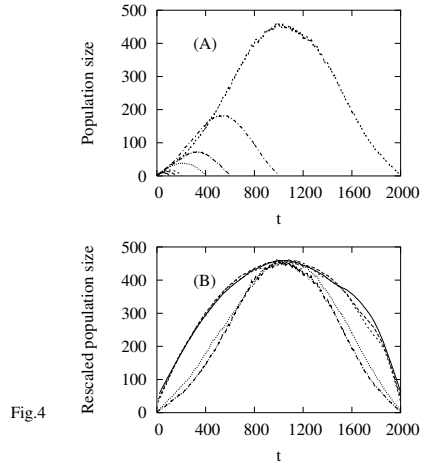}}
%\figspace
\caption{The analysis of the time dependence of the population
size. (A) The average population size of species with a given
lifetime ($p=0.001$, $N=500$). (B) Some data from panel (A)
rescaled (i.e., multiplied by some factors in both directions) in
such a way that the lifetime and the maximal population size
overlap. For species with the lifetime equal to 100, 150, and 200,
the rescaled population sizes nearly overlap. Some deviations from
the overlapping data can be seen for the lifetime 1000 and 2000.}
\label{spsize}
\end{figure}

As our last result, we present the calculation of the average
population size of species of a given lifetime (Fig.\ref{spsize}).
Although all the curves look qualitatively similar, one can notice
a small difference between short- and long-lifetime species. This
difference is better seen on the rescaled plot
(Fig.\ref{spsize}B). This data suggest that population sizes for
species of a lifetime much shorter than the periodicity of
extinctions (which in this case (\cyt{lipowski}) is around 3000)
after rescaling fall into a single curve. For species of a
lifetime comparable or larger than the periodicity of extinctions
the data will deviate from such a universal curve. Although we are
not aware of any palaeontological data of this kind, a comparison
could provide an interesting test of our model.
%%%%%%%%%%%%%%%%%%%%%%%%%%%%%%%%%%%%%%%%%%%%%%%%%%%%%%%%%%%%%%%%%%%%%%%
%%%%%%%%%%%%%%%%%%%%%%%%%%%%%%%%%%%%%%%%%%%%%%%%%%%%%%%%%%%%%%%%%
%%%%%%%%%%%%%%%%%%%%%%%%%%%%%%%%%%%%%%%%%%%%%%%%%%%%%%%%%%%%
%%%%%%%% Fig.3
\subsection{Unique code and the emergence of a multi-species ecosystem}
All living cells use the same code that is responsible for the
transcription of information from DNA to proteins (\cyt{orgel};
\cyt{szathmary}). It suggests that at a certain point of evolution
of life on Earth a replicator that invented this apparently
effective mechanism was able to eliminate replicators of all other
species (if they existed) and establish, at least for a short
time, a single-species ecosystem. Although this process is still
to a large extent mysterious, one expects that subsequent
evolution of these successful replicators leads to their
differentiation and proliferation of species. In such a way the
ecosystem shifted from a single- to multi-species one
(\cyt{lipowski2000}).

It seems to us that the present model might provide some insight
into this problem. As we have already mentioned, the oscillatory
behaviour appears in our model only for the relative update rate
$r<0.27$. When preys reproduce faster ($r>0.27$), a different
behaviour can be seen (\cyt{lipowski}) and the model reaches a
steady state with almost all predators belonging to the same
species with the size $m$ close to 1. Only from time to time a new
species is created with even larger $m$ and a change of the
dominant species might take place. In our opinion, it is possible
that at the very early period of evolution of life on Earth, the ecosystem
resembled the case $r>0.27$. This is because at that time
substrates ('preys') were renewable faster than primitive
replicators ('predators') could use them. If so, every invention
of the increase of the efficiency ('size') could invade the entire
system. In particular, the invention of the coding mechanism could
spread over the entire system. A further evolution increased the
efficiency of predators and that effectively shifted the
(single-species) ecosystem toward
the $r<0.27$ (multi-species, oscillatory) regime.\\
\par
\section{Conclusions}
In the present paper we examined a spatially extended
multi-species prey-predator model. In a certain regime in this
model densities of preys and predators as well as the number of
species show long-term oscillations, even though the dynamics of
the model is not exposed to any external periodic forcing. It
suggests that the oscillatory behaviour of the Earth ecosystem
predicted by Raup and Sepkoski could be simply a natural feature
of its dynamics and not the result of an external factor. Some
predictions of our model such as the lifetime of species or the
time dependence of their population sizes might be testable
against palaeontological data. The prediction that a lifetime of
species depends on the evolutionary stage at which it was created,
that qualitatively agrees with fossil data ~(\cyt{aimiller}),
suggests that a further study of this model would be desirable.

Certainly, our model is based on some restrictive assumptions that
drastically simplifies the complexity of the real ecosystem. We
hope, however, that it includes  some of its important
ingredients: replication, mutation, and competition for resources
(food and space). As an outcome, the model shows that typically
there is no equilibrium-like solution and the ecosystem remains in
an evolutionary cycle. The model does not include geographical
barriers but let us notice that palaeontological data that suggest
the periodicity of mass extinctions are based only on marine
fossils (\cyt{rohde}). More realistic versions should take into
account additional trophic levels, gradual mutations, or sexual
reproduction. One should also notice that the palaeontological
data are mainly at a genus, and not species level. It would be
desirable to check whether the behaviour of our model is in some
sense generic or it is merely a consequence of its specific
assumptions. An interesting possibility in this respect could be
to recast our model in terms of Lotka-Volterra like equations and
use the methodology of adaptive dynamics developed by Dieckmann et
al. (\cyt{DIECKMANN}).

Of course, the real ecosystem was and is exposed to a number of
external factors such impacts of astronomical objects, volcanism
or climate changes. Certainly, they affect the dynamics of an
ecosystem and contribute to the stochasticity of fossil data.
Filtering out these factors and checking whether the main
evolutionary rhythm is indeed set by the ecosystem itself, as
suggested in the present paper, is certainly a difficult task but
maybe worth an effort.
%%%%%%%%%%%%%%%%%%%%%%%%%%%%%%%%%%%%%%%%%%%%%%%%%%%%%%%%%%%%
%%%%%%%%%%%%%%%%%%%%%%%%%%%%%%%%%%%%%%%%%%%%%%%%%%%%%%%%%%%%%
%%%%%%%%%%%%%%%%%%%%%%%%%%%%%%%%%%%%%%%%%%%%%%%%%%%%%%%%%%%%%%%

%Yet another explanation of the observed pattern of extinctions was
%proposed by Stanley\cyt{stanley}. He suggested that relatively
%even and well separated mass extinctions result from delayed
%recovery of an ecosystem. Although convincing, his arguments are
%only of phenomenological nature.

%2) What is a fitness. Is it a size? Does it make sense globally,
%or only within one period?

%3) gradual buildup of an ecosystem. Slow recovery, because
%dominant species slowly increase their size.

\vspace{5mm}
\noindent {\bf Acknowledgements}\\
The research grant 1 P03B 014 27 from KBN is gratefully
acknowledged. We thank Department of Physics of the University of
Aveiro (Portugal) for giving us access to computing facilities.\\
\vspace{5mm}
%%%%%%%%%%%%%%%%%%%%%%%%%%%%%%%%%%%%%%%%%%%%%%%%%%%%%%%%%%%%

%%%%%%%%%%%%%%%%%%%%%%%%%%%%%%%%%%%%%%%%%%%%%%%%%%%%%%%%%%%%%%%%%%%%%%%%%%%%%%%
\end {document}